\def\year{2021}\relax
\documentclass[letterpaper]{article} 
\usepackage{aaai21}  
\usepackage{times}  
\usepackage{helvet} 
\usepackage{color}
\usepackage{courier}  
\usepackage[hyphens]{url}  
\usepackage{graphicx} 
\usepackage{threeparttable}
\usepackage{natbib}  
\usepackage{url}
\usepackage{caption} 
\usepackage[switch]{lineno}  %
\newcommand{\eg}[1]{\textit{e.g.,}}
\newcommand{\ie}[1]{\textit{i.e.,}}

\usepackage{amsmath}
\usepackage{multirow}
\usepackage{multicol}
\usepackage{array, boldline, makecell, booktabs}
\usepackage{amssymb}

\newcommand{\figref}[1]{Fig.~\ref{#1}}
\usepackage{marginnote}

\title{Dual-Octave Convolution for Accelerated Parallel MR Image Reconstruction}

\author{Chun-Mei Feng\textsuperscript{\rm 1}, Zhanyuan Yang\textsuperscript{\rm 3}, Geng Chen\textsuperscript{\rm 4}, Yong Xu\textsuperscript{\rm 1}\textsuperscript{\rm 2}\thanks{Corresponding Author: \textit{Yong Xu}}, Ling Shao\textsuperscript{\rm 4}\\
}
\affiliations{
 	\textsuperscript{\rm 1 }{Shenzhen Key Laboratory of Visual Object Detection and Recognition, Harbin Institute of Technology (Shenzhen), China}\\
 	\textsuperscript{\rm 2 }{Peng Cheng Laboratory, Shenzhen, China}\\
 	\textsuperscript{\rm 3 }{School of Automation Engineering, University of Electronic Science and Technology of China, China}\\
 	\textsuperscript{\rm 4 }{Inception Institute of Artificial Intelligence, Abu Dhabi, UAE} \\
 	{strawberry.feng0304@gmail.com}~~~~~~~~~~~~~{yongxu@ymail.com} 
 }

\begin{document}

\maketitle

\begin{abstract}

\tolerance=1
\emergencystretch=\maxdimen
\hyphenpenalty=10000
\hbadness=10000

Magnetic resonance (MR) image acquisition is an inherently prolonged process, whose acceleration by obtaining multiple undersampled images simultaneously through parallel imaging has always been the subject of research. In this paper, we propose the Dual-Octave Convolution (Dual-OctConv), which is capable of learning multi-scale spatial-frequency features from both real and imaginary components, for fast parallel MR image reconstruction. By reformulating the complex operations using octave convolutions, our model shows a strong ability to capture richer representations of MR images, while at the same time greatly reducing the spatial redundancy. More specifically, the input feature maps and convolutional kernels are first split into two components (i.e., real and imaginary), which are then divided into four groups according to their spatial frequencies. Then, our Dual-OctConv conducts intra-group information updating and inter-group information exchange to aggregate the contextual information across different groups. Our framework provides two appealing benefits: (i) it encourages interactions between real and imaginary components at various spatial frequencies to achieve richer representational capacity, and (ii) it enlarges the receptive field by learning multiple spatial-frequency features of both the real and imaginary components. We evaluate the performance of the proposed model on the acceleration of multi-coil MR image reconstruction. Extensive experiments are conducted on an {in vivo} knee dataset under different undersampling patterns and acceleration factors. The experimental results demonstrate the superiority of our model in accelerated parallel MR image reconstruction.
Our code is available at: github.com/chunmeifeng/Dual-OctConv.
\end{abstract}

\section{Introduction}
Magnetic resonance (MR) imaging has become increasingly popular in radiology and medicine over the past decade, thanks to its advantages of being non-radiative, having a high spatial-resolution, and providing superior soft tissue contrast~\cite{1}. However, a major limitation of MR imaging is that it requires a much longer acquisition time than other imaging techniques, \eg, computed tomography (CT), X-Ray, and ultrasound~\cite{2}. 
Recently, great efforts have been devoted to accelerated MR image reconstruction, which is typically achieved by reconstructing the desired full images from undersampled measured data~\cite{aggarwal2018modl}.


\begin{figure}[t]
\centering
  \includegraphics[width=0.98\linewidth]{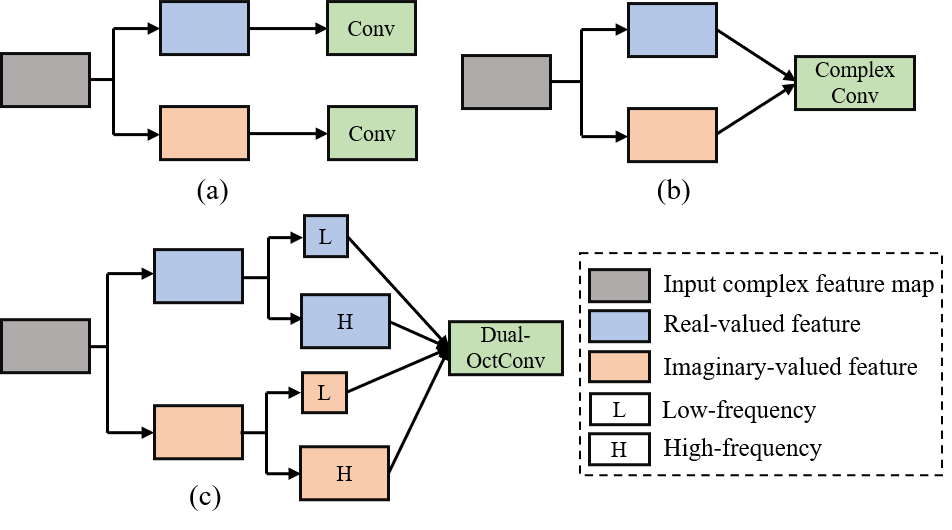}
  \caption{ Previous methods utilize vanilla convolutions (a) to process the real- and imaginary-valued parts of an MR image independently or complex convolutions (b) to jointly deal with the two parts. In contrast, we propose Dual-OctConv, which is a generalization of complex convolutions, to process complex-valued inputs in a multi-frequency space for more effective feature representations.}
  \label{movitation}
\end{figure}

Parallel MR imaging is considered as one of the most important achievements in accelerated MR imaging~\cite{knoll2019deep,wang2017learning}. Most studies (\eg, SENSE~\cite{pruessmann1999sense}, GRAPPA~\cite{griswold2002generalized}, SPIRiT~\cite{lustig2010spirit}) take advantage of spatial sensitivity and gradient coding to reduce the amount of data required for reconstruction, thereby shortening the imaging time. Moreover, compressed sensing (CS) is an important technique for fast MR image reconstruction~\cite{feng2019pca}, which recovers the desired signal from $k$-space data sampled below the Nyquist rate. Typical CS-based approaches adopt a sparsity prior~\cite{liu2019ifr}, low-rank sparse sampling~\cite{he2016accelerated,haldar2016p}, or manifold learning~\cite{nakarmi2017kernel} for reconstruction.


More recently, with the renaissance of deep neural networks, deep learning techniques~\cite{zhou2021cascaded,feng2021coupled}, especially convolutional neural networks (CNNs), have been widely used for parallel MR imaging~\cite{ramani2010parallel,haldar2016p}. Since models are trained offline over large-scale data, only a few extra online samples are required for reconstruction. The model-based unrolling methods~\cite{12,chen2019model} combine mathematical structures (\eg, variational inference, compressed sensing) with deep learning for fast MR image reconstruction. Moreover, extensive approaches~\cite{kwon2017parallel,schlemper2019sigma,schlemper2019data1,sriram2020grappanet,2} propose end-to-end learnable models to remove the aliasing artifacts from images that are reconstructed from undersampled multi-coil $k$-space data. The mapping between a zero-filled $k$-space and fully-sampled MR image is automatically learned by CNNs, requiring no sub-problem division.


Most of the above approaches directly borrow vanilla convolutions used in standard CNNs for $k$-space data in MR image reconstruction. However, vanilla convolutions are designed for real-valued natural images, and cannot deal with complex-valued inputs. To solve this, early studies~\cite{wang} simply discarded the imaginary part or processed the real and imaginary parts independently for real-valued convolutions (see \figref{movitation}(a)). To avoid information loss, complex convolution~\cite{Trabelsi2018DeepCN} has recently been proposed to process complex-valued inputs and encourages information exchange between real and imaginary values (see \figref{movitation}(b)). Though impressive, existing complex convolution operations ignore the intrinsic multi-frequency property of MR images, leading to limited single-scale contextual information and high spatial redundancy in final representations.

To address these limitations, we take a further step towards exploring multi-frequency representation learning in parallel MR image reconstrucion (see~\figref{movitation}(c)). We propose a novel Dual-Octave Convolution (Dual-OctConv), which enables our model to learn multi-frequency representations of multi-coil MR images~\cite{chen2019drop}. Unlike complex convolutions, our Dual-OctConv processes the real (or imaginary) part of MR image features by factorizing it into high- and low-frequency components. The low-frequency component shares information across neighboring locations, and can thus be efficiently processed in low-resolution to enlarge the receptive field and reduce the spatial redundancy. Finally, we combine the features of the real and imaginary parts for reconstruction. Benefiting from Dual-OctConv, our model has a more powerful capability in multi-scale representation learning, and can thus better capture soft tissues (\eg, vascular, muscles) with varying sizes and shapes.



Our main contributions are three-fold: \textbf{First}, we propose multi-frequency feature representations for accelerated parallel MR image reconstruction, and demonstrate their ability to capture multi-scale contextual information. \textbf{Second}, we devise the Dual-OctConv to deal with complex-valued inputs in a multi-frequency representation space, and encourages information exchange across various frequency domains. The Dual-OctConv is a generalization of the standard complex convolution, and endows our model several appealing characteristics (\eg, larger receptive field, higher flexibility, and computationally more efficient).
\textbf{Third}, our model shows significant performance improvements against state-of-the-art algorithms on an \textit{in vivo} knee dataset.

\section{Related Work}
\noindent\textbf{Deep Learning in MR image reconstruction.}
Ever since the pioneering works introducing CNNs for computer vision tasks, such as image classification and face recognition, researchers have made substantial efforts to improve medical and clinical practice using deep learning techniques.~\cite{wang} proposed the first deep learning based MR image reconstruction framework, which learns the mapping between fully-sampled single-coil MR images and their counterpart data reconstructed from a zero-filled undersampled $k$-space. A large number of networks were then developed for MR image reconstruction, especially non-parallel reconstruction~\cite{sun2019deep}. For example, \cite{4} proposed a model-based unrolling method, which formulates the algorithm within a deep neural network, and trained the network with a small amount of data. \cite{han2018deep} employed U-Net to model a domain adaptation structure that removes aliasing artifacts from corrupted images. Similar works have also used deep residual networks~\cite{lee2018deep}, recursive dilated network~\cite{1}, and Generative Adversarial Network~\cite{quan2018compressed, yang2017dagan} to restore high-resolution MR images from undersampled $k$-space data.

In parallel imaging, one representative network is the variational network (VN-Net)~\cite{12}, which combines the mathematical structure of the variational model with deep learning for fast multi-coil MR image reconstruction. Another model-based deep framework~\cite{chen2019model} was designed with a split Bregman iterative algorithm to achieve accurate reconstruction from multi-coil undersampled $k$-space data. To obtain high-fidelity reconstructions, GrappaNet~\cite{sriram2020grappanet} was proposed to combine traditional parallel imaging methods with deep neural networks. Recently, complex-valued representations have demonstrated superiority in processing complex-valued inputs~\cite{Trabelsi2018DeepCN}. For example,~\citet{2} applied complex convolutions to jointly process real and imaginary values for comprehensive feature representations. In contrast, our approach represents complex-valued input features in a multi-frequency space. The Dual-OctConv, proposed for processing such multi-frequency data, can capture richer contextual knowledge, leading to significant improvement in performance.

\begin{figure}[t]
\centering
  \includegraphics[width=0.48\textwidth]{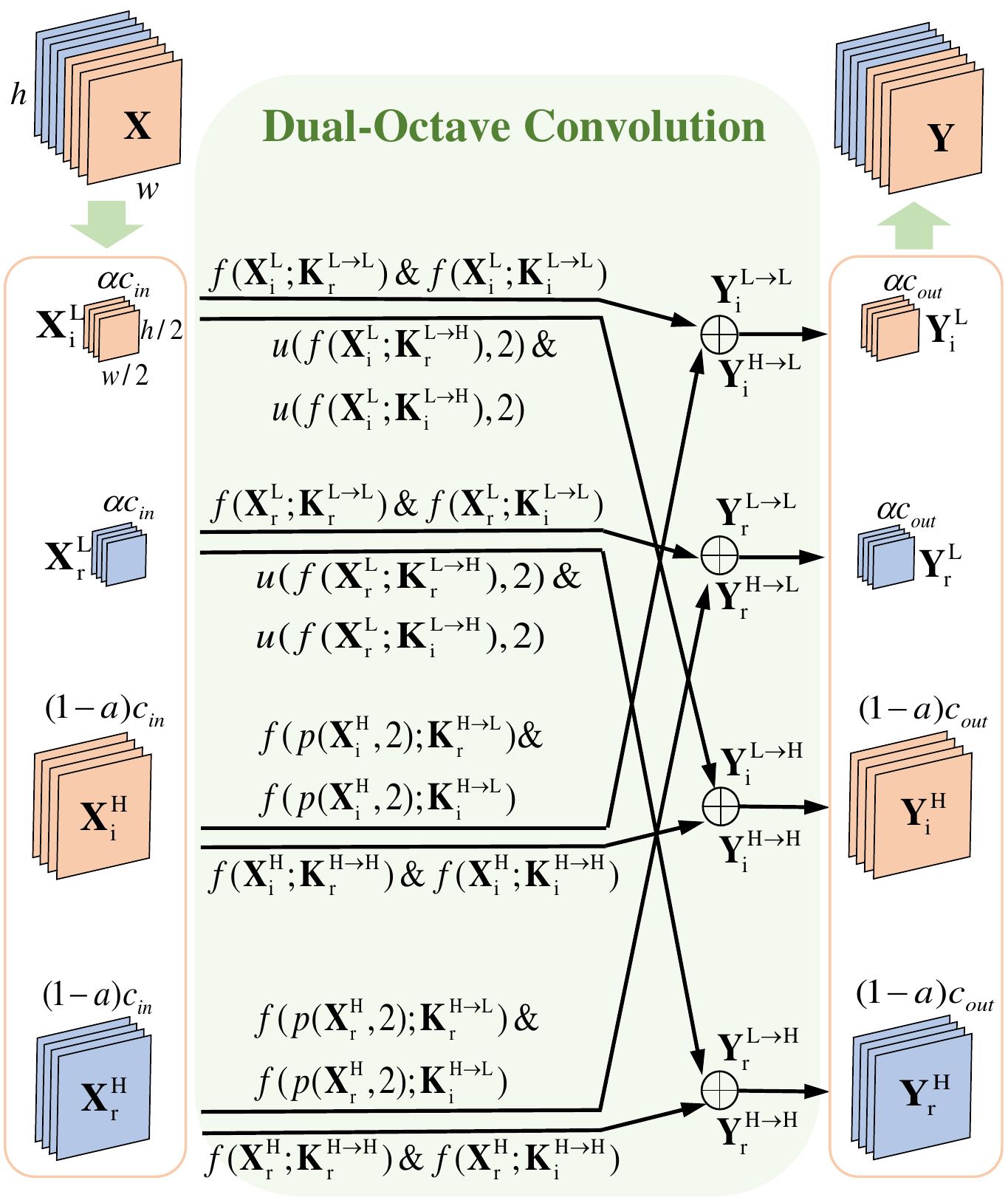}
  \caption{ Detailed design of our Dual-OctConv. $\mathbf{X}\!\in\!\mathbb{C}^{c\!\times\!h\!\times\!w}$ represents the input complex-valued feature maps, and $\mathbf{Y}\!\in\!\mathbb{C}^{c\!\times\!h\!\times\!w}$ indicates the corresponding output feature maps, modulated by the Dual-OctConv. $u$ and $p$ denote the upsampling and average pooling operations, respectively. Please see Eq.~\eqref{eq:5} for more details.}
  \label{(a)}
\end{figure}

\noindent\textbf{Multi-Scale Representation Learning.}
Multi-scale information has proven effective in various computer vision tasks (\eg, image classification, object detection, semantic segmentation). Several strategies have been proposed for multi-scale representation learning, yielding significant performance improvement in a number of tasks. For example, \citet{ke2017multigrid} proposed a multi-grid network to propagate and integrate information across multiple scales for image classification. Multi-scale information has also been proven effective in restoring image details for image enhancement~\cite{nah2017deep,ren2016single,li2018multi}. In addition, various 
well-known techniques (\eg, FPN~\cite{lin2017feature} and PSP~\cite{zhao2017pyramid}) have been proposed for learning multi-scale representations in object detection and segmentation tasks~\cite{zhou2020motion,zhou2020matnet}. Recently, the Octave convolution~\cite{chen2019drop} was proposed to learn multi-scale features based on the spatial frequency model, greatly improving performance in natural image and video recognition. 

In this work, we demonstrate the appealing properties of the Octave convolution for accelerated parallel MR image reconstruction, which helps to capture multi-scale information from multiple spatial-frequency features. Based on this, we propose a novel Dual-OctConv for accelerated parallel MR image reconstruction, which enables our model to capture details of vasculatures and tissues with varying sizes and shapes, yielding high-fidelity reconstructions.

\section{Methodology}
\subsection{Problem Formulation}
MR scanners acquire $k$-space data through the receiver coils and then utilize an inverse multidimensional Fourier transform to obtain the final MR images.
In parallel imaging, multiple receiver coils are used to simultaneously acquire $k$-space data from the target under scanning.


Let $\mathbf A\!=\!\mathbf M\!\mathbf F\!\in\!\mathbb{C}^{M\times N}$ denote the undersampled Fourier encoding matrix, where $\mathbf F$ is the multidimensional Fourier transform, and $\mathbf M$ is an undersampled mask operator.
In parallel imaging, the same mask is used for all coils.
The undersampled $k$-space data from each coil can be expressed as
\begin{equation}\small
\mathbf y_i = \mathbf A (\mathbf{S}_i\mathbf x), \label{XX}
\end{equation}
where $i=1,2,...,c$, with $c$ denoting the number of coils,
$\mathbf x\!\in\!\mathbb{C}^{N\times 1}$ is the ground truth MR image,
$\mathbf y_i\!\in\!\mathbb{C}^{M\times 1}(M\!<\!<\!N)$ is the undersampled $k$-space data for the $i$-th coil,
and $\mathbf{S}_i$ is a complex-valued diagonal matrix encoding the sensitivity map of the $i$-th coil.
The coil sensitivities modulate the $k$-space data, which is measured by each coil.
The coil configuration and interactions with the anatomical structures under scanning can affect coil sensitivities, so $\mathbf{S}_i$ changes across different scans.
In addition, the obtained image will contain aliasing artifacts, if the inverse Fourier transform is directly applied to undersampled $k$-space data.

We can reconstruct $\hat{\mathbf x}$ with prior knowledge of its properties, which is formulated as the following  problem:
\begin{equation}\small
	\hat{\mathbf x}=\mathop{\arg \min}_{\mathbf x}\sum_{i=1}^c \  \| {\mathbf y_i - \mathbf A (\mathbf{S}_i\mathbf x) } \|^2_2+ \ \lambda \Psi(\mathbf x),
	\label{eq:object_function}
\end{equation}
where $\Psi$ is a regularization function and $\lambda$ controls the trade-off between the two terms.

The problem presented in Eq.~\eqref{eq:object_function} can be effectively resolved using CNNs, which avoids time-consuming numerical optimization and the need of a coil sensitivity map.
During training, we update the network weights as follows: 
\begin{equation}\small
	\hat{\boldsymbol{\theta}}= \mathop{\min}_{\boldsymbol{\theta}} \frac{1}{N}\sum_{n=1}^{N} \| {\mathbf{x'}(n) - f_{\boldsymbol{\theta}}( \mathbf{y'}(n) )} \|_1,
\label{eq:theta}
\end{equation}
where $\mathbf{y'}(n)$ is the $n$-th multi-channel image obtained from the zero-filled $k$-space data, $\mathbf{x'}(n)$ is the $n$-th ground truth multi-channel image, $N$ is the total number of training samples, and $f_{\boldsymbol{\theta}}(\cdot)$ is an end-to-end mapping function parameterized by $\boldsymbol{\theta}$, which contains a large number of adjustable network weights.
Training with Eq.~\eqref{eq:theta} can reconstruct the expected MR images, but the original information of the data acquired in the $k$-space cannot be well preserved.
If we incorporate the undersampled $k$-space data into the data fidelity at the training stage, the network can yield improved reconstruction results.
For this purpose, we add the data fidelity units in our network, as in \cite{2}.
After the network is trained, we obtain a set of optimal parameters $\boldsymbol{\hat \theta}$ for the reconstruction of multi-channel image, and predict the multi-channel image via
$\mathbf{\hat{x}'} = f_{\hat{\boldsymbol{\theta}}}(\mathbf{y'})$.
Finally, we use an adaptive coil combination method \cite{2} to obtain the expected MR image from $\mathbf{\hat{x}'}$.
\begin{figure}[t]
\centering
  \includegraphics[width=0.48\textwidth]{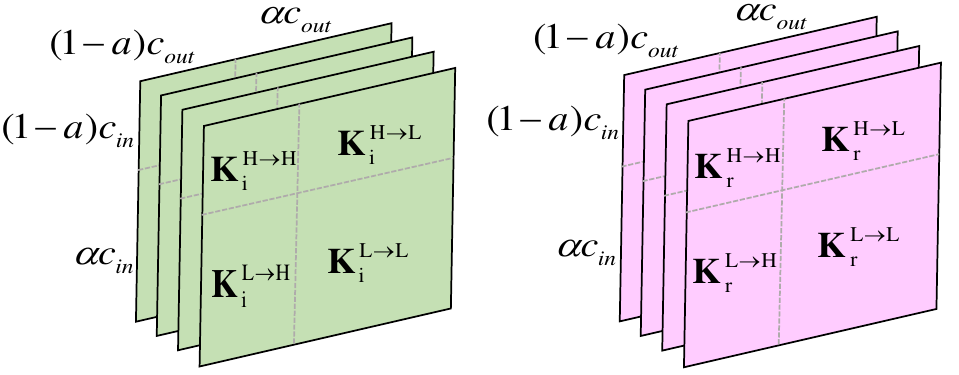}
  \caption{The Dual-OctConv kernels. Green and pink squares denotes the imaginary and real kernels, respectively.}
  \label{(b)}
\end{figure}

\begin{figure*}[h]
\centering
  \includegraphics[width=0.98\textwidth]{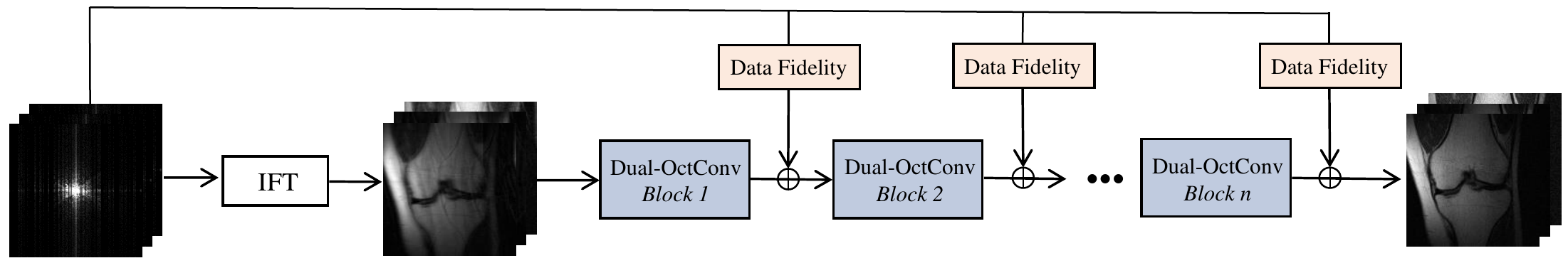}
  \caption{ Architecture of our network for parallel MR image reconstruction. The input is a set of zero-filled multi-coil $k$-space measurements, while the output is the reconstructed multi-channel MR image. $\rm IFT$ represents the 2D inverse Fourier transform.}
  \label{flowchart}
\end{figure*}

\subsection{Dual-Octave Convolution}
To obtain rich multi-scale context information, we first represent the multi-channel input with complex filters, and then decompose it into low and high spatial frequency parts.
Let $\mathbf X \!\in\!\mathbb{C}^{c\times h\times w}$ be the the complex feature maps with $c$, $h$, and $w$ denoting the number of channels, height, and width, respectively.
As illustrated in~\figref{(a)}, we convolve $\mathbf X = \mathbf X_{\rm r}+i\mathbf X_{\rm i}$ with a complex filter matrix $\mathbf K = \mathbf K_{\rm r}+i\mathbf K_{\rm i}$. Mathematically, we have
\begin{equation}\small
{\left[ \begin{array}{ccc}
\Re(\mathbf K* \mathbf X)\\
\Im(\mathbf K* \mathbf X)
\end{array} 
\right ]}= {\left[ \begin{array}{ccc}
\mathbf K_{\rm r} & -\mathbf K_{\rm i}\\
\mathbf K_{\rm i} &   \mathbf K_{\rm r}
\end{array} 
\right ]} *
{
\left[ \begin{array}{ccc}
\mathbf X_{\rm r} \\
\mathbf X_{\rm i} 
\end{array} 
\right ]},
\end{equation}
where the matrices $\mathbf K_{\rm r}$ and $\mathbf K_{\rm i}$ represent real and imaginary kernels, and vectors $\mathbf X_{\rm r}$ and $\mathbf X_{\rm i}$ represent real and imaginary feature maps.
Note that all the kernels and feature maps are expressed by real matrices since the complex arithmetics are simulated by real-valued entities.

To effectively fuse the real and imaginary parts of data (\ie, $\mathbf X_{\rm r}$ and $\mathbf X_{\rm i}$), we split them into low and high spatial frequency groups $\mathbf X_{\rm r} = \left\{ \mathbf X_{\rm r}^{\rm L}, \mathbf X_{\rm r}^{\rm H} \right\}$, $\mathbf X_{\rm i} = \left\{ {\mathbf X_{\rm i}^{\rm L}, \mathbf X_{\rm i}^{\rm H}} \right\}$, where $\mathbf X^{\rm H}\!\in\!\mathbb{C} ^{(1-\alpha) c \times h \times w}$ capture the high-frequency fine details of the data and $\mathbf X^{\rm L}\!\in\!\mathbb{C} ^{\alpha c \times 0.5h \times 0.5w}$ determine the low-frequency image contrast.
Here, $\alpha\!\in\![0,1]$ controls the ratio of channels that are allocated to low-frequency and high-frequency feature maps. 
Note that the Dual-OctConv will turn into the standard complex convolution~\cite{Trabelsi2018DeepCN} when $\alpha\!=\!0$. 
As shown in~\figref{(b)}, the complex filter matrix is further expressed as $\mathbf K_{\rm r}^{\rm H} = \left[\mathbf K_{\rm r}^{{\rm H}\rightarrow {\rm L}},\mathbf K_{\rm r}^{{\rm H}\rightarrow {\rm H}} \right]$, $\mathbf K_{\rm i}^{\rm H} = \left[\mathbf K_{\rm i} ^{{\rm H}\rightarrow {\rm L}},\mathbf K_{\rm i}^{{\rm H}\rightarrow {\rm H}} \right]$, $\mathbf K_{\rm r}^{\rm L} = \left[\mathbf K_{\rm r}^{{\rm L}\rightarrow {\rm H}},\mathbf K_{\rm r}^{{\rm L}\rightarrow {\rm L}} \right]$, $\mathbf K_{\rm i}^{\rm L} = \left[\mathbf K_{\rm i}^{{\rm L}\rightarrow {\rm H}},\mathbf K_{\rm i}^{{\rm L}\rightarrow {\rm L}} \right]$ to convolve with $\mathbf X_{\rm r}^{\rm L}$, $\mathbf X_{\rm i}^{\rm L}$, $\mathbf X_{\rm r}^{\rm H}$ and $\mathbf X_{\rm i}^{\rm H}$.
We then have
\begin{equation}\small
\begin{split}
     \mathbf Y_{\rm r}^{\rm L} =&f(\mathbf X_{\rm r}^{\rm L};\mathbf K_{\rm r}^{{\rm L}\rightarrow {\rm L}})+u(f(\mathbf X_{\rm r}^{\rm L};\mathbf K_{\rm r}^{{\rm L}\rightarrow H}),2) \\
     &+f(\mathbf X_{\rm r}^{\rm H};\mathbf K_{\rm i}^{{\rm L}\rightarrow {\rm L}})+u(f(\mathbf X_{\rm r}^{\rm L};\mathbf K_{\rm i}^{{\rm L}\rightarrow {\rm H}}),2),\\
     \mathbf Y_{\rm i}^{\rm L} =&f(\mathbf X_{\rm i}^{\rm L};\mathbf K_{\rm r}^{{\rm L}\rightarrow {\rm L}})+u(f(\mathbf X_{\rm i}^{\rm L};\mathbf K_{\rm r}^{{\rm L}\rightarrow {\rm H}}),2)\\
     &-f(\mathbf X_{\rm i}^{\rm {\rm H}};\mathbf K_{\rm i}^{{\rm L}\rightarrow {\rm L}})-u(f(\mathbf X_{\rm i}^{\rm L};\mathbf K_{\rm i}^{{\rm L}\rightarrow {\rm H}}),2),\\
     \mathbf Y_{\rm r}^{\rm H} =&f(\mathbf X_{\rm r}^{\rm H};\mathbf K_{\rm r}^{{\rm H}\rightarrow {\rm H}})+f(p(\mathbf X_{\rm r}^{\rm H},2);\mathbf K_{\rm r}^{{\rm H}\rightarrow {\rm L}}))\\
     &+f(\mathbf X_{\rm r}^{\rm H};\mathbf K_{\rm i}^{{\rm H}\rightarrow {\rm H}})+f(p(\mathbf X_{\rm r}^{\rm H},2);\mathbf K_{\rm i}^{{\rm H}\rightarrow {\rm L}})),\\
     \mathbf Y_{\rm i}^{\rm H} =&f(\mathbf X_{\rm i}^{\rm H};\mathbf K_{\rm r}^{{\rm H}\rightarrow {\rm H}})+f(p(\mathbf X_{\rm i}^{\rm H},2);\mathbf K_{\rm r}^{{\rm H}\rightarrow {\rm L}}))\\
     &-f(\mathbf X_{\rm i}^{\rm H};\mathbf K_{\rm i}^{{\rm H}\rightarrow {\rm H}})-f(p(\mathbf X_{\rm i}^{\rm H},2);\mathbf K_{\rm i}^{{\rm H}\rightarrow {\rm L}})),\\
\end{split}
\label{eq:5}
\end{equation}
where $f(\mathbf X;\mathbf K)$ denotes the convolution with parameters $\mathbf K$, $u(\mathbf X,k)$ denotes the upsampling operation with a factor of $k$ via nearest interpolation, $p(\mathbf X,k)$ denotes the average pooling with kernel size $k \times k$, and $c(\cdot)$ denotes the concatenation operation.
The real and imaginary parts are fused with the operations $\left\{ {\rm L}\rightarrow {\rm L},{\rm H}\rightarrow {\rm H} \right\}$ and $\left\{ {\rm H}\rightarrow {\rm L},{\rm L}\rightarrow {\rm H} \right\}$, which correspond to the information updating and exchanging between high- and low-frequency feature maps.
Therefore, our Dual-OctConv is able to enlarge the receptive fields of the low-frequency feature maps both in the real and imaginary parts. To put this into perspective, after convolving the low-frequency feature maps of the real and imaginary parts ( $\mathbf X_{\rm r}^{\rm L}$, $\mathbf X_{\rm i}^{\rm L}$) with $k \times k$ complex convolution kernels, the receptive fields of both  achieve a 2$\times$ enlargement compared to the vanilla convolution.
Thus, our Dual-OctConv has a strong ability to capture rich context information at different scales.
Finally, we compute the final output feature maps as
\begin{equation}\small
\mathbf Y = u(c(\mathbf Y_{\rm r}^{\rm L}\cdot \mathbf Y_{\rm i}^{\rm L}))+c(\mathbf Y_{\rm r}^{\rm H} \cdot \mathbf Y_{\rm i}^{\rm H}).
\end{equation}
In a nutshell, we first split the real and imaginary parts of input feature map $\mathbf X$ into low and high spatial frequency components.
Then, all these components are convolved with the corresponding complex filter to obtain the new components, where the information is effectively fused.
Finally, these components are concatenated to obtain the final output feature maps $\mathbf Y$.

\subsection{Detailed Network Architecture}
Based on the proposed Dual-OctConv, we design an effective deep learning model for accelerated parallel MR image reconstruction. As shown in~\figref{flowchart}, our network consists of ten Dual-OctConv blocks, each of which is comprised of five Dual-OctConv layers which are organized in a residual form.
Our network can be trained in an end-to-end manner with the training data from $\mathbf{y'}(n)$ and $\mathbf{x'}(n)$.
The input is an undersampled multi-coil $k$-space measurement and the output is the reconstructed MR image.
We first transform the $k$-space data to obtain aliased multi-channel images, before feeding them into the following Dual-OctConv blocks.
Following~\cite{2}, we add a data fidelity unit between consecutive blocks to preserve the original $k$-space information during training.
In each layer except the last one, we use ReLU as the activation function.

\section{Experiments}

\begin{table*}[t]
 \small
 \centering
 \setlength\tabcolsep{2.2pt}
 \renewcommand\arraystretch{1.3}
 \begin{tabular}{r||cc|cc|cc|cc|cc|cc|cc|cc}
  \hlineB{2.5}
  & \multicolumn{4}{c|}{1D Uniform}  &  \multicolumn{4}{c|}{1D Cartesian} & \multicolumn{4}{c|}{2D Random} & \multicolumn{4}{c}{2D Radial} \\ \cline{2-17} 
  & \multicolumn{2}{c|}{3x} & \multicolumn{2}{c|}{5x}  & \multicolumn{2}{c|}{3x} & \multicolumn{2}{c|}{5x} & \multicolumn{2}{c|}{3x} & \multicolumn{2}{c|}{5x} & \multicolumn{2}{c|}{4x} & \multicolumn{2}{c}{6x} \\ \cline{2-17}
       & PSNR & SSIM & PSNR & SSIM & PSNR & SSIM & PSNR & SSIM & PSNR & SSIM & PSNR & SSIM & PSNR & SSIM & PSNR & SSIM \\ \hline\hline
  Zero-filing & 24.406 & 0.676  & 23.579 & 0.655 & 25.922 & 0.726  & 24.550 & 0.685  & 30.540 & 0.827  & 27.078 & 0.750  & 31.026 & 0.826 &28.107 & 0.766 \\
  SPIRiT &29.385 & 0.700  &28.300 & 0.676   &32.310 & 0.801  &31.222 & 0.782 &32.179 & 0.786  &32.258 & 0.812   &30.308 &0.720  &29.061 &0.702 \\
  L1-SPIRiT &29.815 & 0.847  &27.353 & 0.788   &33.346 & 0.887  &30.912 & 0.837 &38.597 & 0.937  &34.071 & 0.887   &37.004  &0.919 &34.149 &0.881\\
  VN-Net &35.436 & 0.907  &32.730 & 0.858   &36.364 & 0.912  &33.236 & 0.866 &38.409 & \textcolor[rgb]{1.00,0.00,0.00}{0.956}  &35.734 & 0.923   & 37.956 & 0.930 &34.609  &0.907 \\
  ComplexMRI &34.989 & 0.909  &32.803 & 0.873   &35.957 & 0.916  &34.126 & 0.876 &39.563 & 0.946  &37.315 & 0.908   &38.098  &0.933 &35.768 &0.904\\ \hline
  Dual-OctConv &\textcolor[rgb]{1.00,0.00,0.00}{36.243} & \textcolor[rgb]{1.00,0.00,0.00}{0.919}  &\textcolor[rgb]{1.00,0.00,0.00}{34.128} & \textcolor[rgb]{1.00,0.00,0.00}{0.885}   &\textcolor[rgb]{1.00,0.00,0.00}{37.029} & \textcolor[rgb]{1.00,0.00,0.00}{0.923}  &\textcolor[rgb]{1.00,0.00,0.00}{34.944} & \textcolor[rgb]{1.00,0.00,0.00}{0.884} &\textcolor[rgb]{1.00,0.00,0.00}{39.964} & 0.948  &\textcolor[rgb]{1.00,0.00,0.00}{38.279} & {\color{red}0.930}   &{\color{red}38.607}  &{\color{red}0.935} &{\color{red}36.584} &{\color{red}0.908} \\ \hline

 \end{tabular}
 \caption{ Quantitative comparison of state-of-the-art methods under different undersampling patterns. Best results are marked in red. }
 \label{t1}
\end{table*}

\begin{figure*}[!htb]
\centering
  \includegraphics[width=0.99\textwidth]{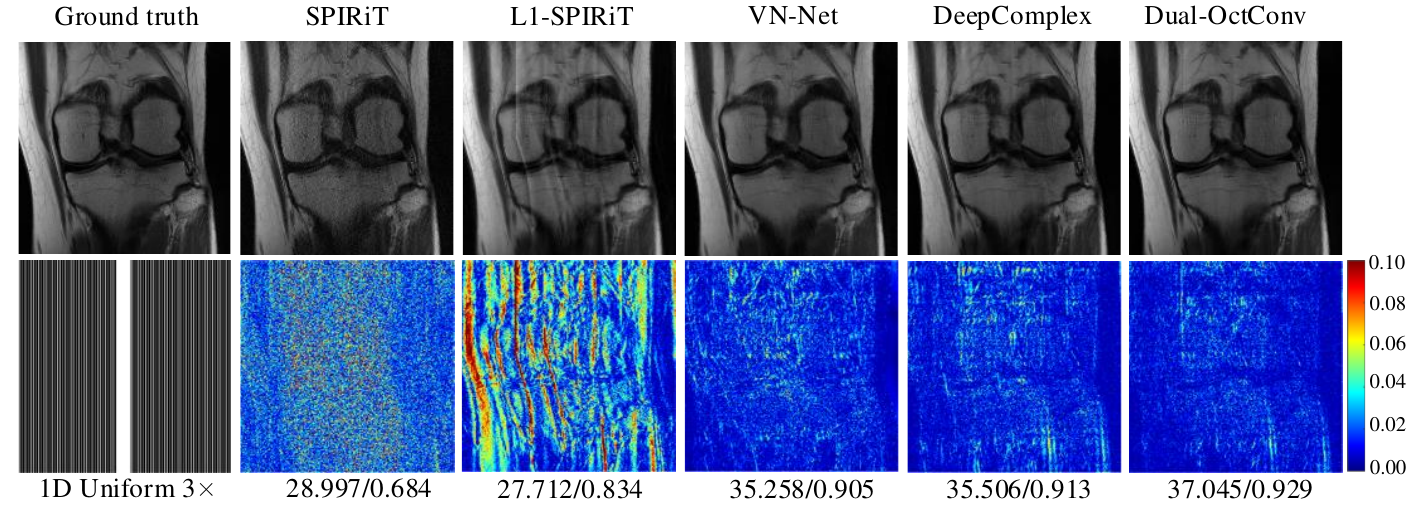}
  \caption{ Comparison of different methods in terms of reconstruction accuracy, with {1D undersampling patterns} and a {3$\times$} acceleration rate. Reconstruction results and error maps are presented with corresponding quantitative measurements in PSNR/SSIM.}
  \label{1D}
\end{figure*}

\begin{figure*}[!htb]
\centering
  \includegraphics[width=0.99\textwidth]{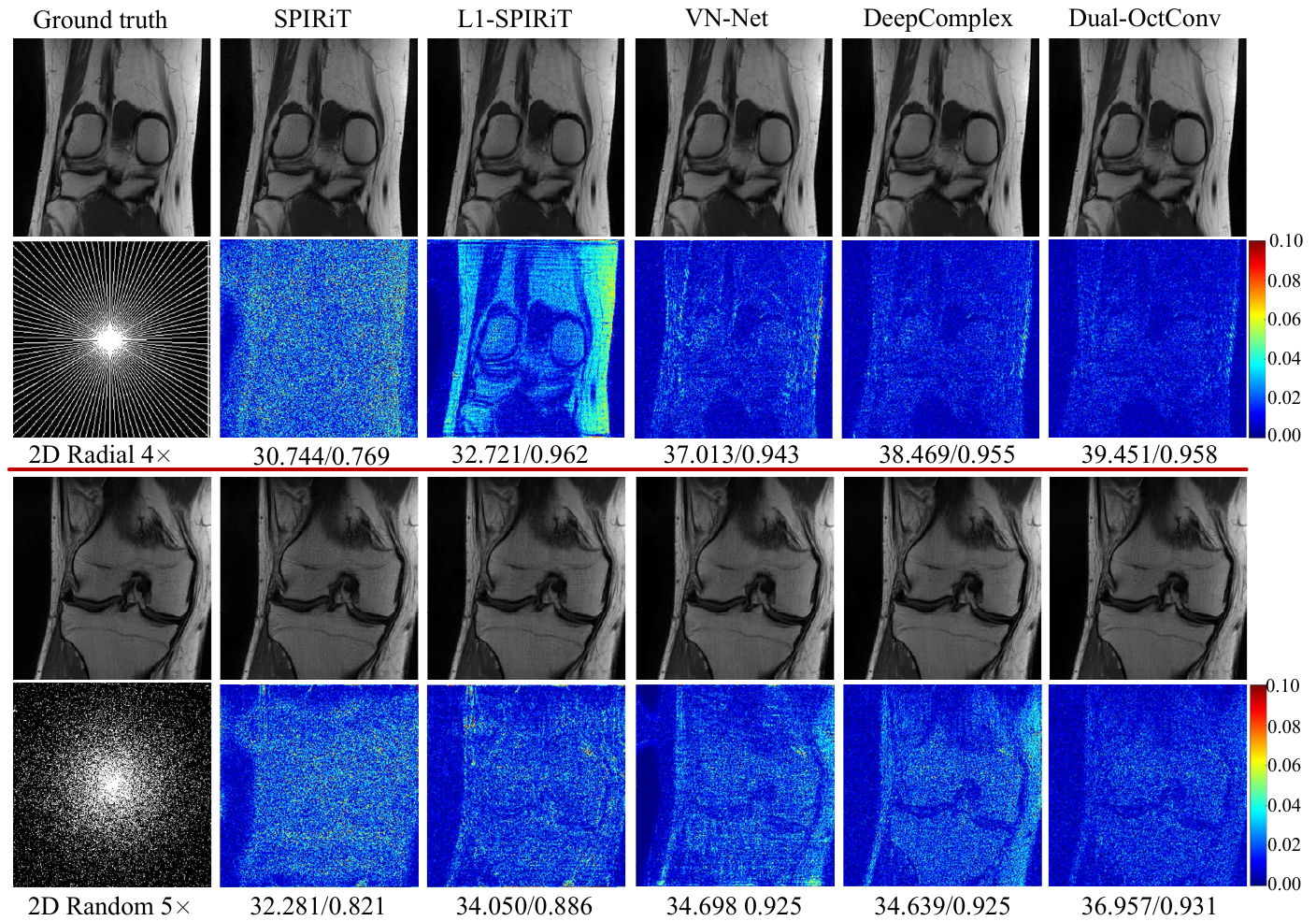}
  \caption{Comparison of different methods in terms of reconstruction accuracy, with {2D undersampling patterns} and {4$\times$} and {5$\times$} acceleration rates, respectively. Reconstruction results and error maps are presented with corresponding quantitative measurements in PSNR/SSIM.}
  \label{2D}
\end{figure*}

\subsection{Datasets}
We use the \textit{in vivo} multi-coil fully-sampled MR knee dataset that is acquired using a clinical 3T Siemens Magnetom Skyra scanner with a sequence called `Coronal Spin Density Weighted without Fat Suppression'\!~\cite{12}. 
The imaging protocol is detailed as follows: 15-channel knee coil, matrix size $320\!\times320\!\times\!20$, TR=2750 ms, TE=27 ms, and in-plane resolution = 0.49$\times$0.44 mm$^2$.
There are 20 subjects in total: 5 female/15 male, age 15-76, and BMI 20.46-32.94.
We randomly select 14 patients for training, 3 for validation, and 3 for testing.

The pre-defined undersampling masks are used to obtain the undersampled measurements.
In our experiments, we adopt four different $k$-space undersampling patterns, including 1D uniform, 1D Cartesian, 2D random, and 2D radial.
Examples of the undersampling patterns are illustrated in~\figref{1D} and~\figref{2D}.
For 1D uniform, 1D Cartesian, and 2D random masks, the acceleration rate is set to 3 and 5.
For the 2D radial mask, 4$\times$ and 6$\times$ accelerations are adopted.

\subsection{Implementation Details}
We implement our model using Tensorflow 1.14 and perform experiments using an NVIDIA 1080Ti GPU with a 11GB memory. Following~\cite{2}, we initialize the magnitude and phase of the complex parameters using Rayleigh and uniform distributions, respectively.
The network is trained using the Adam optimizer \cite{2} with initial learning rate 0.001 and weight decay 0.95.
The batch size is set to 4 and convolutional kernel size is set to $3\times 3$. Each complex convolutional layer has 64 feature maps, except for the last layer, which is determined by the concatenated real and imaginary channels of the data. The spatial frequency ratio $\alpha$ is set to $0.125$ by default.

To demonstrate its effectiveness, we compare our Dual-OctConv with a number of state-of-the-art parallel MR imaging approaches, including traditional methods (SPIRiT \cite{lustig2010spirit} and L1-SPIRiT \cite{murphy2010clinically}) as well as CNN-based methods (VN-Net \cite{12} and ComplexMRI \cite{2}). All these methods are trained on the same dataset with their default settings. For CNN-based methods, we re-trained them according to the specifications with TensorFlow, using their default parameter settings.

\subsection{Quantitative Evaluation}
We use peak signal-to-noise ratio (PSNR) and structural similarity index measure (SSIM)~\cite{2} for quantitative evaluation. Table~\ref{t1} reports the average PSNR and SSIM results with respect to different undersampling patterns and acceleration factors. As can be seen, our Dual-OctConv obtains consistent performance improvements against the baseline methods, across various settings.
Additionally, we observe that the undersampling patterns greatly affect the quality of reconstruction. For instance, the 2D sampling masks generally outperform the 1D masks. Another important observation is that the reconstruction becomes more difficult when the acceleration rate increases.



In particular, our model significantly outperforms previous methods under extremely challenging settings (\ie, 2D masks with 5$\times$ and 6$\times$ acceleration). This can be attributed to the powerful capability of our Dual-OctConv in aggregating rich contextual information of real and imaginary data. Moreover, we see that our model show strong robustness under various undersampling patterns and acceleration rates.


\subsection{Qualitative Evaluation}
For qualitative analysis, we first show the reconstructed images and corresponding error maps for 1D uniform with a 3$\times$ acceleration rate in~\figref{1D}. In general, our method provides the best-quality reconstructed images and significantly reduces prediction errors. In contrast, the baseline methods yield large prediction errors and show unsatisfactory performance. In particular, compared with the CNN-based method, the traditional methods show obvious errors, such as SPIRiT and L1-SPIRiT.

We next examine the results for 2D radial masks with a 4$\times$ acceleration rate and 2D random masks with a 5$\times$ acceleration rate. As shown in~\figref{2D}, CNN-based methods significantly improve the results with less errors and clearer structures, in comparison with SPIRiT and L1-SPIRiT. In particular, our Dual-OctConv produces higher-quality images with clear details and minimum artifacts. The superior performance is owed to the fact that Dual-OctConv can effectively aggregate the information of various spatial frequencies present in the real and imaginary parts of an MR image.

\subsection{Ablation Studies}

\begin{figure}[t]
\centering
  \includegraphics[width=0.47\textwidth]{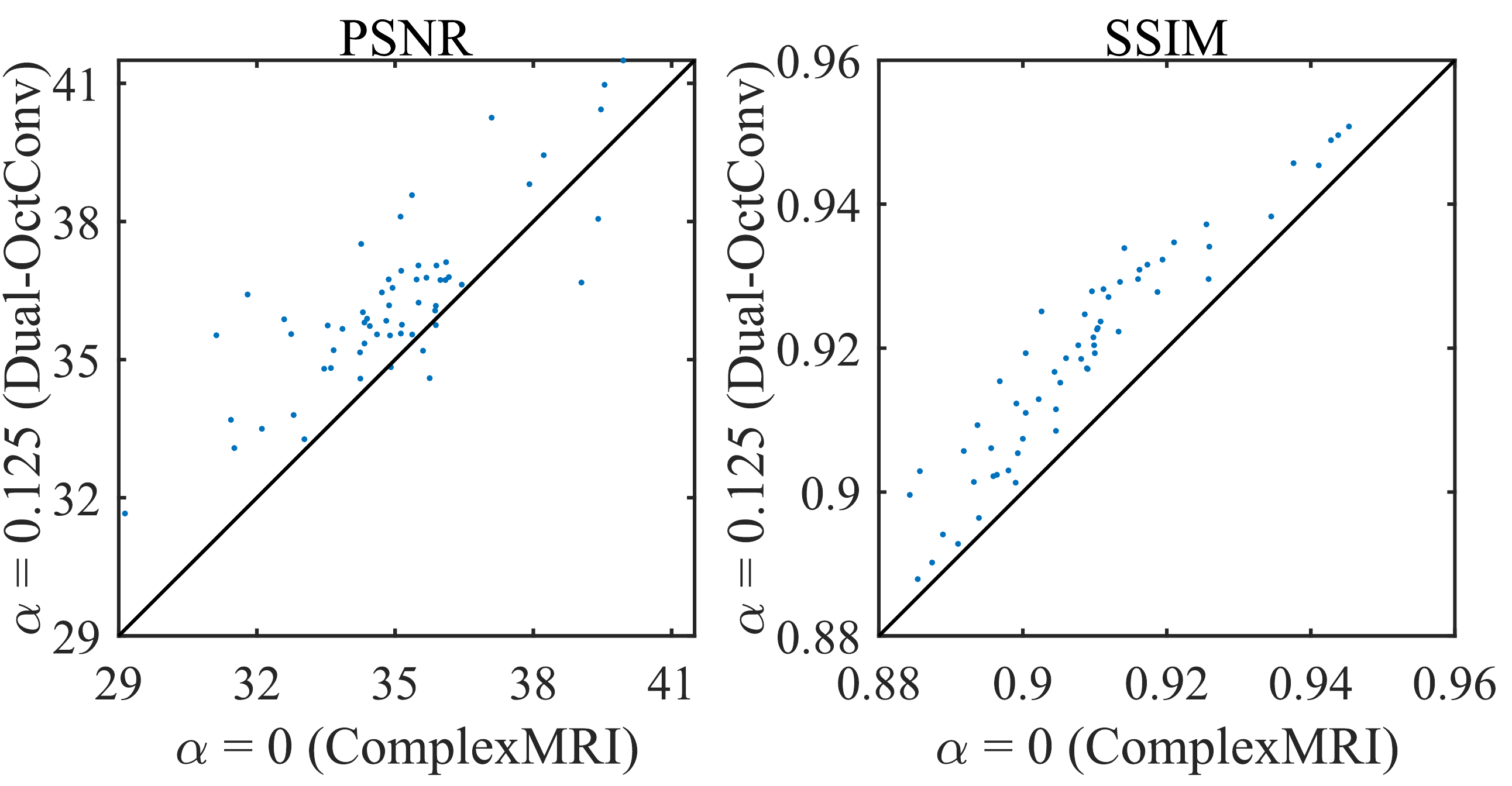}
  \caption{Quantitative comparison of Dual-OctConv and the baseline model ($\alpha\!=\!0$) in terms of PSNR and SSIM. Note that ratio $\alpha\!=\!0$ is equivalent to the complexMRI model.}
  \label{withoutoct}
\end{figure}

Firstly, we study the effects of the proposed Dual-OctConv. For comparison, we build a baseline model by setting $\alpha\!=\!0$, which turns the Dual-OctConv into a standard complex convolution. We conduct experiments on the test set with 60 complex-valued images under the uniform undersampling mask with a 3$\times$ acceleration rate. As illustrated in~\figref{withoutoct}, our Dual-OctConv significantly outperforms the baseline model, especially in terms of SSIM. This reveals the superiority of the Dual-OctConv in improving the reconstruction.

\begin{figure}[t]
\centering
  \includegraphics[width=0.47\textwidth]{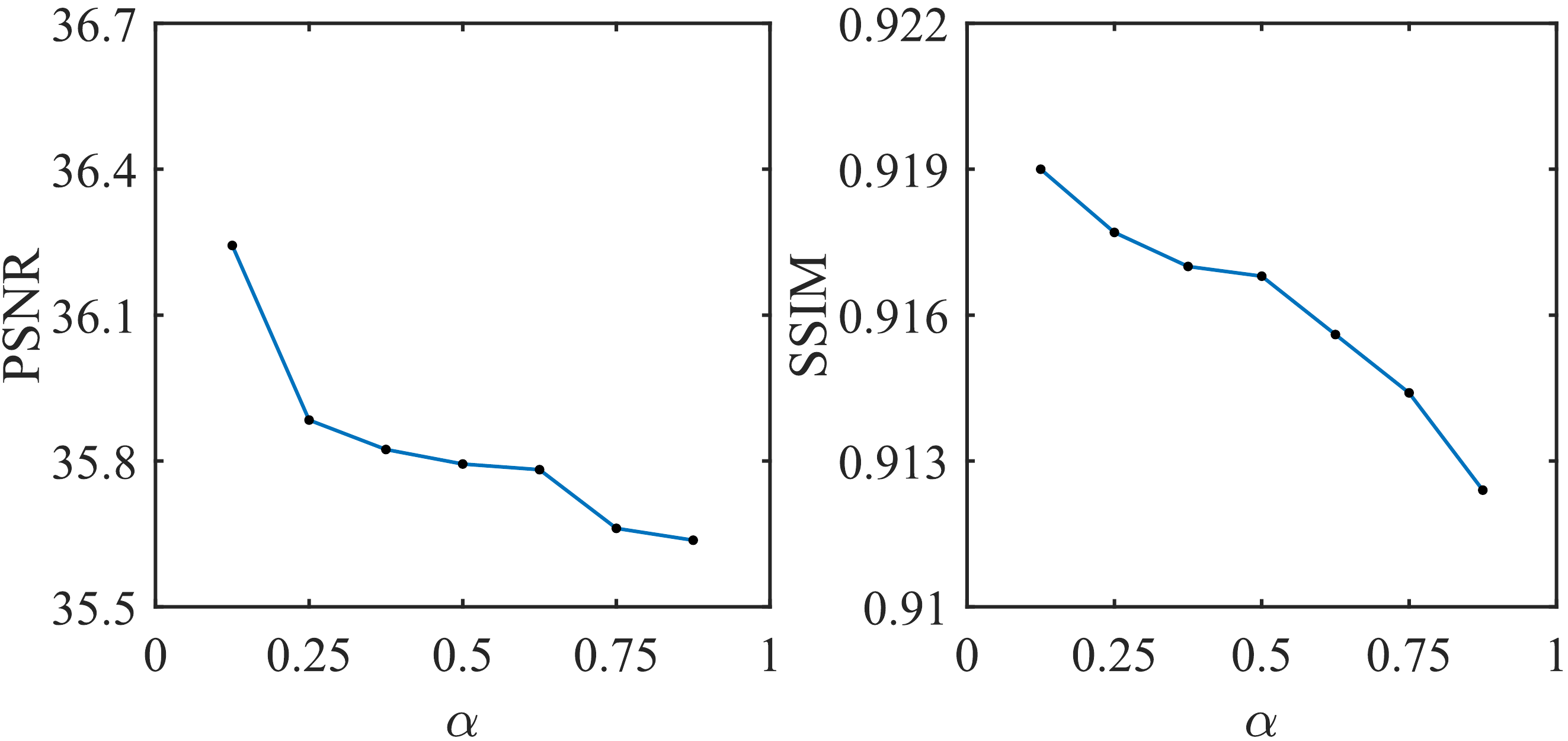}
  \caption{ Analysis of spatial frequency ratio ($\alpha$) in terms of PSNR and SSIM.}
  \label{alpha}
\end{figure}

Secondly, we investigate the influence of the spatial frequency ratio $\alpha$ for reconstruction. The ratio determines the receptive fields in both the real and imaginary parts, and also influences the fusion of these parts at multiple spatial frequencies. As shown in~\figref{alpha}, our model shows the best PSNR and SSIM scores at $\alpha\!=\!0.125$, which means that 12.5\% of the channels in the real and imaginary parts are reduced to a low spatial frequency. When $\alpha$ becomes larger, the performance quickly degrades due to severe information loss induced by over-large ratios.


\begin{figure}[t]
	\centering
	\includegraphics[width=0.47\textwidth]{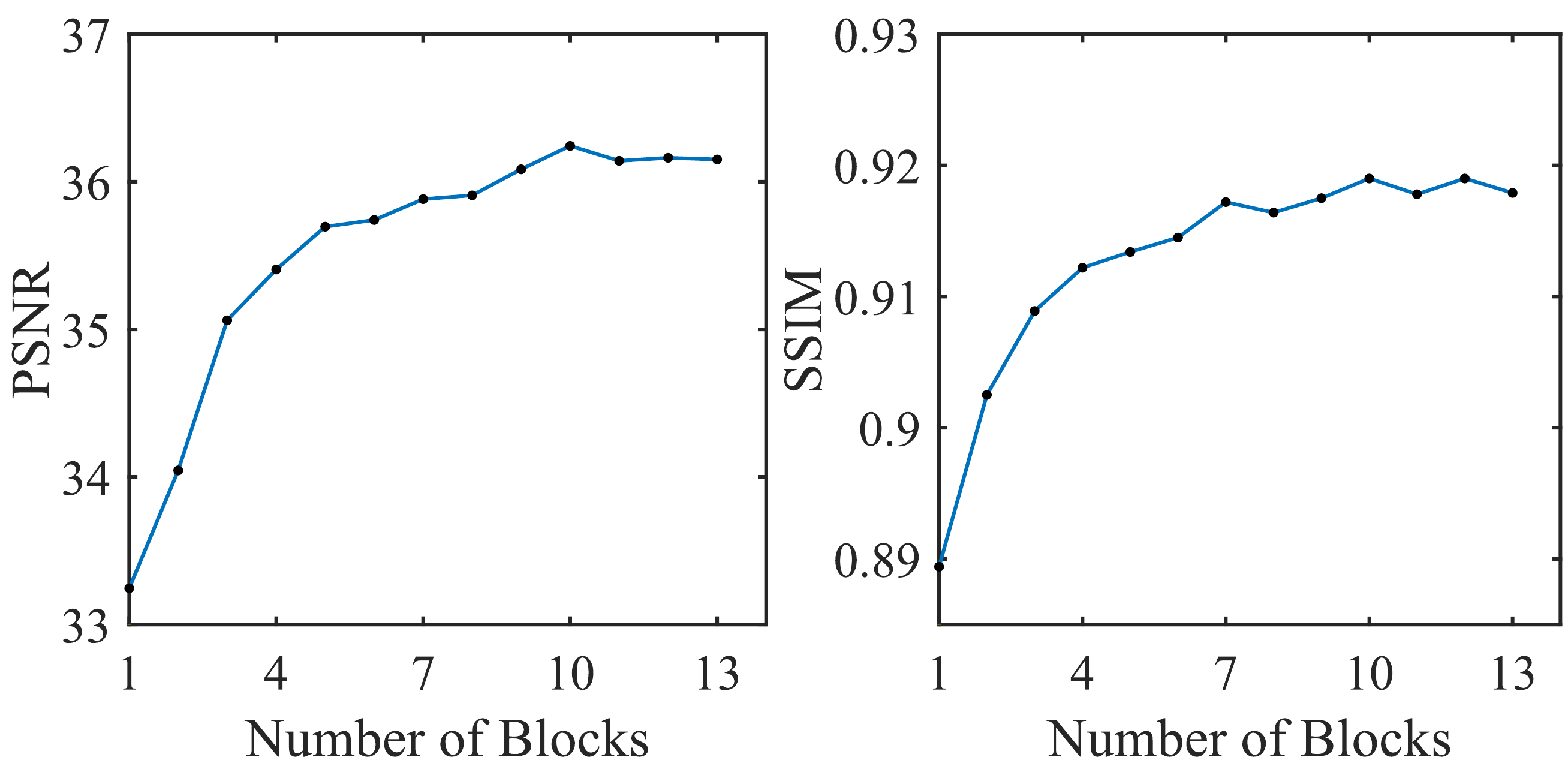}
	\caption{ Performance comparison of our network with respect to the number of Dual-OctConv blocks.}
	\label{block}
\end{figure}

\begin{figure}[t]
\centering
  \includegraphics[width=0.47\textwidth]{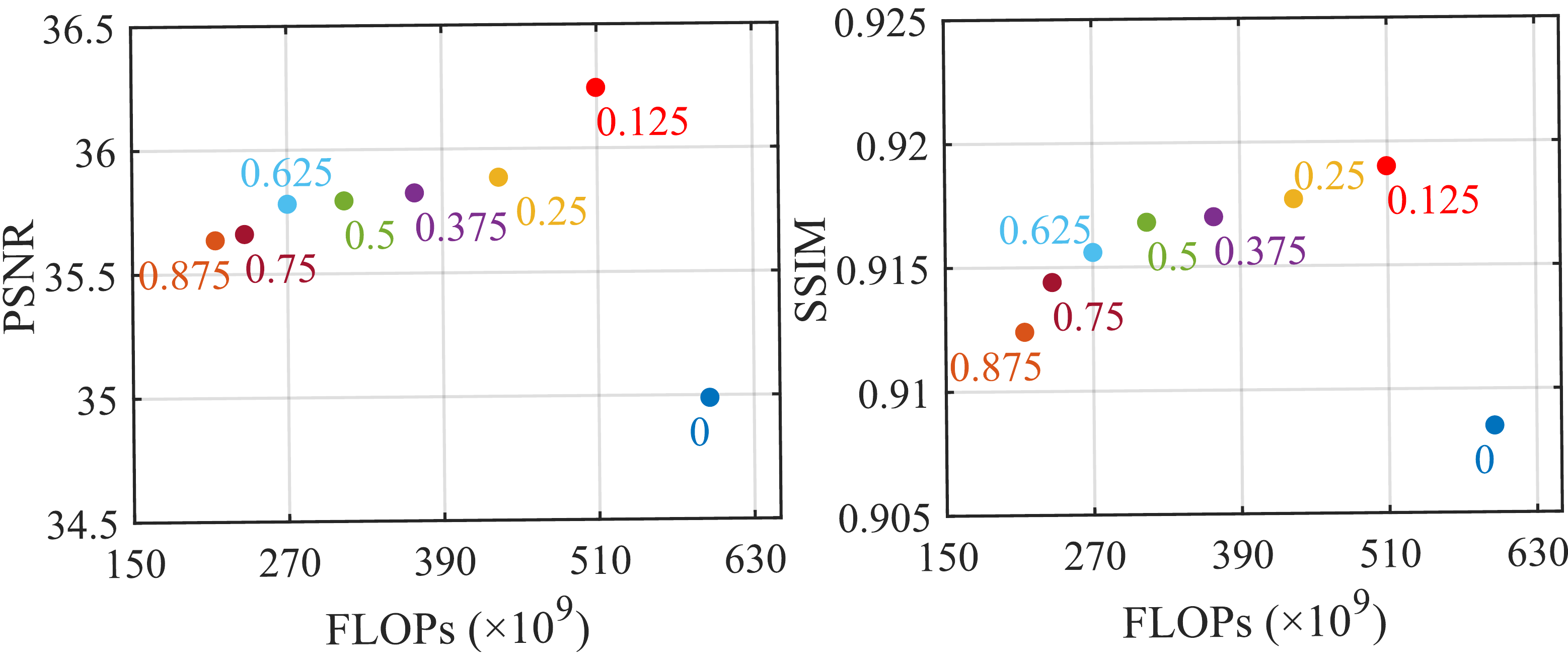}
  \caption{ FLOPs analysis with respect to spatial frequency ratios ($\alpha$). The number below each point is the value of $\alpha$. We see that, under varioous settings ($0\!<\!\alpha\!<\!1$), our Dual-OctConv is always more efficient and accurate than the baseline model ($\alpha\!=\!0$).}
  \label{flops}
\end{figure}

The number of network parameters increases as the number of blocks ($bn$) increases. Therefore, it is necessary to choose an appropriate number of blocks to ensure that the network structure reaches the highest reconstruction accuracy without inducing higher computational and memory requirements.
Herein, we carry out various experiments using different numbers of blocks. The results are presented in~\figref{block}.
As can be seen from the curves, our model can successfully reconstruct the MR images at $bn\!=\!4$, and the reconstruction accuracy reaches the highest at $bn\!=\!10$.

Finally, we study the FLOPs of Dual-OctConv with respect to different $\alpha$ in~\figref{flops}.
The number below each point is the value of $\alpha$, and $\alpha\!=\!0$ refers to the baseline model.
As can be observed, a small $\alpha$ leads to improved performance with a higher FLOPs. Moreover, compared with the baseline model (\ie, $\alpha\!=\!0$), our model consistently shows better performance with much lower FLOPs.



\section{Conclusion}
In this work, we focus on spatial frequency feature expression in complex-valued data for parallel MR image reconstruction.
For this purpose, we propose a novel Dual-OctConv operation to deal with the real and imaginary components at multiple spatial frequencies. By convolving the feature maps of both the real and imaginary components under different spatial resolutions,
our Dual-OctConv is able to reconstruct higher-quality images with significantly reduced artifacts.
We conduct extensive experiments on an $in~vivo$ knee dataset under different settings of undersampling patterns and acceleration rates, and the results demonstrate the advantages of our model against state-of-the-art methods in accelerated MR image reconstruction.



\bibliographystyle{aaai}
\bibliography{bibliography}

\end{document}